\documentclass[aps,prb,twocolumn,showpacs,superscriptaddress]{revtex4-1}
\usepackage{graphicx}
\begin{document}

\title{The atomic structure of the $\sqrt{3} \times \sqrt{3}$ phase of silicene on Ag(111)}

\author{Seymur Cahangirov}
\affiliation{Nano-Bio Spectroscopy Group and ETSF Scientific Development Centre, Departamento de F\' isica de Materiales, Universidad del Pa\' is Vasco, CSIC-UPV/EHU-MPC and DIPC, Avenida de Tolosa 72, E-20018 San Sebastian, Spain}

\author{Veli Ongun \" Oz\c celik}
\affiliation{Institute of Material Science and Nanotechnology, Bilkent University, Ankara 06800, Turkey}

\author{Lede Xian}
\affiliation{Nano-Bio Spectroscopy Group and ETSF Scientific Development Centre, Departamento de F\' isica de Materiales, Universidad del Pa\' is Vasco, CSIC-UPV/EHU-MPC and DIPC, Avenida de Tolosa 72, E-20018 San Sebastian, Spain}

\author{Jose Avila}
\affiliation{Synchrotron SOLEIL, L'Orme des Merisiers, Saint Aubin-BP 48, 91192 Gif sur Yvette Cedex, France}

\author{Suyeon Cho}
\affiliation{Synchrotron SOLEIL, L'Orme des Merisiers, Saint Aubin-BP 48, 91192 Gif sur Yvette Cedex, France}

\author{Mar\' ia C. Asensio}
\affiliation{Synchrotron SOLEIL, L'Orme des Merisiers, Saint Aubin-BP 48, 91192 Gif sur Yvette Cedex, France}

\author{Salim Ciraci} \email{ciraci@fen.bilkent.edu.tr}
\affiliation{Department of Physics, Bilkent University, Ankara 06800, Turkey}

\author{Angel Rubio} \email{angel.rubio@ehu.es}
\affiliation{Nano-Bio Spectroscopy Group and ETSF Scientific Development Centre, Departamento de F\' isica de Materiales, Universidad del Pa\' is Vasco, CSIC-UPV/EHU-MPC and DIPC, Avenida de Tolosa 72, E-20018 San Sebastian, Spain}

\date{\today}

\begin{abstract}
The growth of the $\sqrt{3} \times \sqrt{3}$ reconstructed silicene on Ag substrate has been frequently observed in experiments while its atomic structure and formation mechanism is poorly understood. Here by first-principles calculations we show that $\sqrt{3} \times \sqrt{3}$ reconstructed silicene is constituted by dumbbell units of Si atoms arranged in a honeycomb pattern. Our model shows excellent agreement with the experimentally reported lattice constant and STM image. We propose a new mechanism for explaining the spontaneous and consequential formation of $\sqrt{3} \times \sqrt{3}$ structures from $3 \times 3$ structures on Ag substrate. We show that the $\sqrt{3} \times \sqrt{3}$ reconstruction is mainly determined by the interaction between Si atoms and have weak influence from Ag substrate. The proposed mechanism opens the path to understanding of multilayer silicon.
\end{abstract}


\maketitle

\section{Introduction}

Silicene, a monolayer of silicon atoms arranged in a honeycomb lattice, received an enormous interest for being a candidate two-dimensional material that could bring the exotic electronic structure of graphene\cite{geim} to the well-developed silicon-based technology \cite{seymurprl,spinhall,vogtprl,diboride,ezawa,feng,scirep,chen,induced,spontane,padova,ir111,nature,naturecom,lelay}. Single-layer free-standing silicene has been predicted to be stable \cite{seymurprl} and was experimentally synthesized on Ag (111) substrates \cite{vogtprl,feng,scirep,chen,induced,spontane,padova}. Unlike graphene, free standing silicene does not have a planar structure but attains its stability by minor buckling, whereby alternating Si atoms are located in different planes to attain $sp^3$-like bonding \cite{seymurprl,takeda,durgun}. Despite this buckling, free-standing silicene preserves linearly crossing bands at the Fermi level that leads to Dirac Fermion behavior of its electrons. Hence, the need of unraveling the exotic electronic structure of silicene \cite{spinhall,ezawa,naturecom} and its remarkable integration in the well-established silicon technology have placed silicene at the forefront of intensive theoretical and experimental research. 

Most of the experimental work have been concentrated on the growth of silicene on Ag (111) substrate with few but yet important exceptions \cite{diboride,ir111}. An open debate is if the interaction between silicene and Ag substrate is weak or if it is strong enough to destroy the linearly crossing bands \cite{chen,induced,absence,hybridization}. The structural properties of silicene on Ag surface have proved to be intricate and strongly dependent on the growth conditions \cite{mess}. Nevertheless, there are two structures, which have been unanimously supported by theoretical and experimental reports \cite{vogtprl,feng,scirep,chen,induced,spontane,padova}. The first one is the so called "flower pattern", which can be described as the 3$\times$3 superstructure with respect to silicene lattice commensurately matched with 4$\times$4 supercell of Ag (111) surface. The flower pattern has been shown to be dictated by the interaction of silicene with Ag substrate and is formed by three protruding Si atoms arranged in hexagons \cite{vogtprl,induced,hybridization}. The atomic structure of this configuration is well understood by theoretical calculations and scanning tunneling microscopy (STM) measurements, while there is still a debate on the origin of the linear bands observed by the angle resolved photoemission (ARPES) experiments \cite{vogtprl,induced,hybridization,avila} even if the proposal that they come from Si-Ag hybridized state is gaining consensus \cite{hybridization,avila}.

The other silicene superstructure frequently observed on Ag (111) substrate has a $\sqrt{3} \times \sqrt{3}$ periodicity with respect to silicene lattice \cite{feng,scirep,chen,spontane,padova}. ARPES measurements have clearly shown that the $\Gamma$K direction of 1$\times$1 silicene is aligned with the $\Gamma$K direction of the Ag (111) surface in both 3$\times$3 and $\sqrt{3} \times \sqrt{3}$ superstructures \cite{padova}. However, it is not clear how the $\sqrt{3} \times \sqrt{3}$ reconstructed silicene phase can be commensurately matched with the lattice of Ag (111) keeping this alignment. In fact, so far, there is no model that could explain the origin of $\sqrt{3} \times \sqrt{3}$ reconstruction or the mechanism behind the compression of silicene lattice by $\sim$~5~$\%$ as observed in STM measurements. In particular, the model proposed by Chen \textit{et al.} considers a variation in the buckling pattern of regular 1$\times$1 silicene by pushing down one of the upper sublattice Si atom in every $\sqrt{3} \times \sqrt{3}$ supercell below the level of the lower sublattice atoms \cite{chen}. In the freestanding case this configuration is less energetically favorable compared to the regular silicene and it remains so until the structure is squeezed more than 5~$\%$, which is below the experimental lattice constant. Therefore, this structure is energetically unfavorable and there is no reason for this particular reconstruction to occur. Moreover, this structure can not be stabilized by the Ag substrate since there is no lattice matching between the two.

Here, by first-principles calculations, we unveil the microscopic structure of freestanding $\sqrt{3} \times \sqrt{3}$ silicene. In our model Si atoms form dumbbell (DB) geometries arranged in the honeycomb lattice of $\sqrt{3} \times \sqrt{3}$ supercell. This arrangement excellently reproduces the STM images reported for $\sqrt{3} \times \sqrt{3}$ silicene. It also explains the spontaneous $\sim$5~$\%$ compression of the lattice constant. Furthermore, we introduce a growth model where first $3 \times 3$ reconstructed silicene is formed on Ag substrate and then it is gradually transformed to $\sqrt{3} \times \sqrt{3}$ silicene which becomes incommensurate with Ag substrate. For this novel phase, the cohesive energy per Si atom is higher than that of free-standing silicene and the phonon dispersions are positive over the whole Brillouin zone (BZ), confirming their structural stability even if they are freestanding. Comprehensively, our findings show that the $\sqrt{3} \times \sqrt{3}$ phase of silicene diverges from graphene-like band structure and provide a remarkable new playground for outstanding applications ranging from photovoltaics to molecular electronics.

\section{Methods}

Ab-initio Density Functional Theory (DFT) calculations were carried out using the projector-augmented wave (PAW) pseudopotential method \cite{paw} as implemented in the VASP software \cite{vasp}. The generalized gradient approximation (GGA) in the Perdew-Burke-Ernzerhof form \cite{pbe} was used to include the exchange-correlation interactions. A plane wave energy cutoff of 300 eV was used. The vacuum spacing between image surfaces due to the periodic boundary condition is larger than 12.5~\AA. The forces on the relaxed atoms were converged to less than 10$^{-3}$~eV/\AA. In the ionic relaxation calculations, the Brillouin zone was sampled by (11$\times$11$\times$1) k-points. The Tersoff-Hamann model was used for the simulation of STM images \cite{stm}. Ab-initio molecular dynamics simulations were carried out at 500 K for 2 picoseconds. A semi-empirical dispersion potential is used to include the van der Walls interaction \cite{grimme}.

\section{Results and Discussions}

We first present the results of our first-principles molecular dynamics (MD) simulation describing the formation mechanism of the DB units \cite{kaltsas,ongun}, as building blocks of the stable $\sqrt{3} \times \sqrt{3}$ phases as well as their relation to the well-known 3$\times$3 phase of silicene on Ag(111). In Fig.~1(a) we present a pictorial summary of the MD simulations. We start from perfect silicene 3$\times$3 layer and place additional Si atom far away from it. Then, Si ad atoms are attached to silicene by dangling bonds and at an intermediate stage they form bridge bonds with two-second neighbor Si atoms of silicene thereby increasing the coordination number of these Si atoms from three to four. In search for tetrahedral orientation, these four bonds then force the atoms to move towards directions shown in the middle panel of Fig.~1(a). As a result, the new Si ad atom sits $\sim$1.38~\AA~above the top site of silicene while at the same time pushing down the Si atom just below it by the same amount. The same also happens when we start with 3$\times$3 reconstructed silicene on Ag(111) substrate. The DB formation is an exothermic process and occurs spontaneously without need to overcome any kind of barrier. Consequently, we have found that the formation of individual DBs is inevitable in a medium comprising free Si atoms and silicene \cite{ongun}.

\begin{figure}
\includegraphics[width=8.5cm]{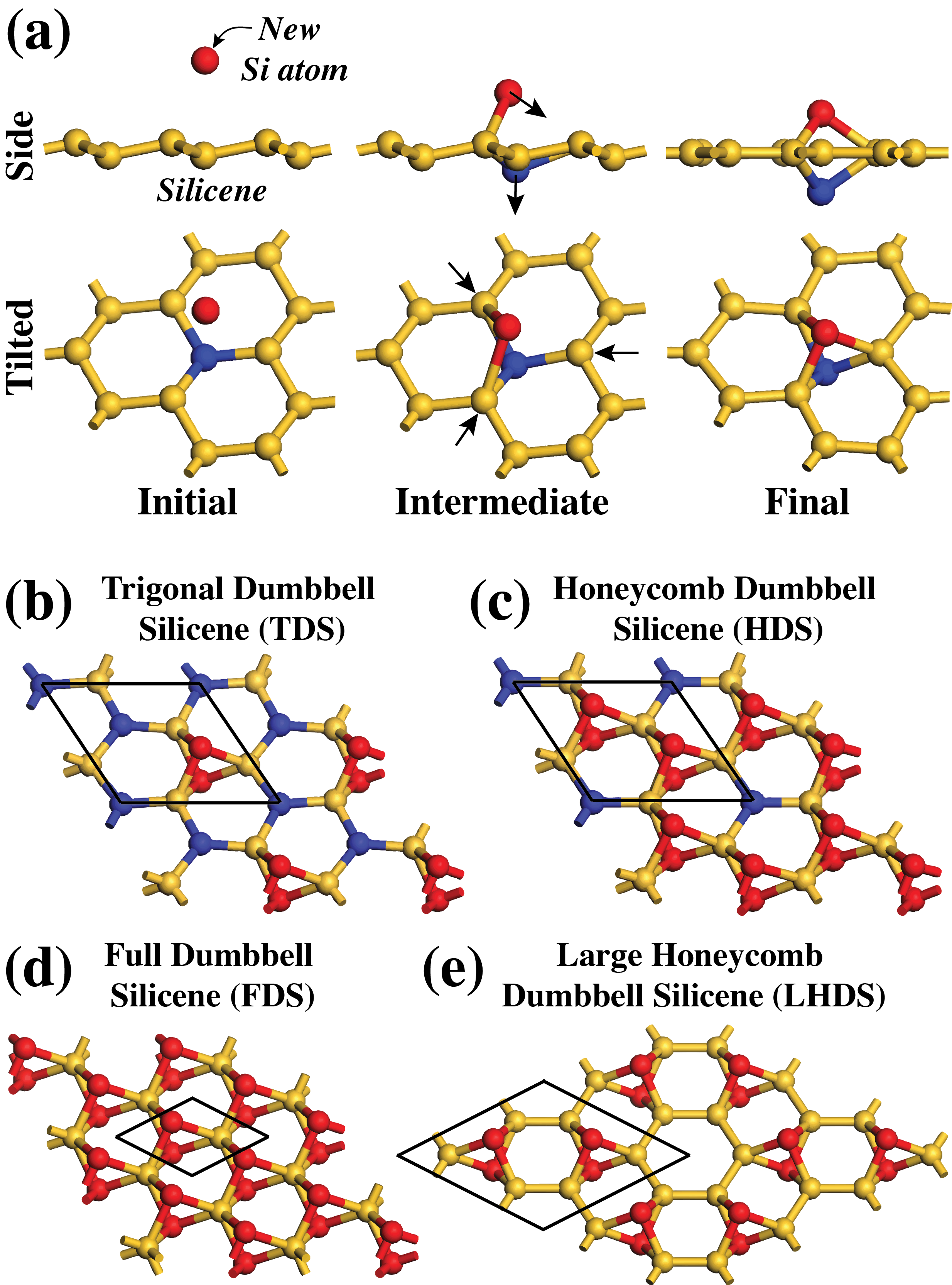}
\caption{(a) Formation of the dumbbell building block units starting from buckled silicene. Atomic structure of (b) $\sqrt{3} \times \sqrt{3}$ trigonal dumbbell silicene (TDS), (c) $\sqrt{3} \times \sqrt{3}$ honeycomb dumbbell silicene (HDS), (d) $1 \times 1$ full dumbbell silicene (FDS) and (e) $2 \times 2$ large honeycomb dumbbell silicene (LHDS). The unitcells are delineated by solid black lines. Atoms having different environment are represented by balls having different colors.}
\label{fig1}
\end{figure}

Our calculations show that when a single DB unit is placed in an n$\times$n unitcell the cohesive energy per Si atom is maximized when n=$\sqrt{3}$ and decreases monotonically for n$\ge$2. We refer to the structure having single DB unit in $\sqrt{3} \times \sqrt{3}$ unitcell as trigonal dumbbell silicene (TDS) due to the trigonal lattice formed by DB atoms, as shown in Fig.~1(b) \cite{kaltsas}. As seen in Table I, TDS is energetically more favorable than free-standing silicene, for which the cohesive energy per Si atom is 3.958 eV/atom \cite{kaltsas,ongun}. Interestingly, the cohesive energy per Si atom is further increased when another DB unit is created in the $\sqrt{3} \times \sqrt{3}$ unitcell of TDS. We refer to this new structure as honeycomb dumbbell silicene (HDS) due to the honeycomb structure formed by two DB units in $\sqrt{3} \times \sqrt{3}$ unitcell (see Fig.~1(c)). Adding another DB unit in the $\sqrt{3} \times \sqrt{3}$ unitcell of HDS results in a 1$\times$1 structure composed of DB atoms connected by sixfold coordinated Si atoms (see Fig.~1(d)). This structure, that we refer to as full dumbbell silicene (FDS), has a cohesive energy of 3.973~eV/atom, which is less than that of TDS and HDS.

We should emphasize that it is the interplay between two competing effects that makes HDS the most favorable $\sqrt{3} \times \sqrt{3}$ structure. While formation of new DBs and thus new bonds increases the cohesive energy, the increase in the coordination number beyond four decreases it. As seen in Fig.~1, the coordination number of yellow atoms in TDS structure is four while in HDS it is five. Apparently, the formation of a new DB and hence new bonds compensates the energy required to form the peculiar five-fold coordination. However, it fails to compensate the six-fold coordination of Si atoms forming the middle atomic layer of FDS. This arguments led us to investigate another DB structure that has even larger cohesive energy per atom compared to HDS. This structure has two DB units arranged in a honeycomb lattice in a 2$\times$2 unitcell. Here the packing of DB units is dense compared to TDS but sparse compared to HDS. In this structure, the honeycomb lattice formed by dumbbell units is larger compared to the one formed in HDS, hence we refer to this structure as large honeycomb dumbbell silicene (LHDS). As seen in Fig.~1(e), the maximum coordination of Si atoms in the LHDS is four. Since there are more DB units in LHDS compared to TDS and no hypervalent Si atoms as in HDS, the cohesive energy per atom of free-standing LHDS is higher than both TDS and HDS.

\begin{table}
\begin{center}
\begin{tabular}{cccccc}
\hline
\space Structures \space & \begin{tabular}{@{}c@{}} $\sqrt{3} \times \sqrt{3}$ \\ \space\space lattice (\AA) \space\space \end{tabular} & \multicolumn{2}{c}{\begin{tabular}{@{}c@{}} Energy per \\ \space\space atom (eV/atom) \space\space \end{tabular}} & \multicolumn{2}{c}{\begin{tabular}{@{}c@{}} Energy per \\ \space\space area (eV/\AA$^2$) \space\space \end{tabular}} \\
\hline
\space & \space &  Free  &  On Ag  &  Free &  On Ag  \\
3$\times$3  & 6.67 &  3.850 & 4.877 & 0.598 & 0.759\\
TDS & 6.52 & 4.013 & 4.663 & 0.764 & 0.887 \\
LHDS & 6.43 & 4.161 & 4.483 & 0.871 & 0.938 \\
HDS & 6.38 & 4.018 & 4.471 & 0.912 & 1.014 \\
\hline
\end{tabular}
\end{center}
\caption{Lattice constants and cohesive energies of the four different phases of silicene studied in the present work. For comparison the lattice constant is given in terms of the length corresponding to $\sqrt{3} \times \sqrt{3}$ cell, although it is not the unitcell of 3$\times$3 and LHDS structures. Cohesive energies are calculated both in the absence and presence of the Ag substrate and presented both in terms of energy per atom and energy per area.} \label{table1}
\end{table}

We performed a stringent stability check of the structures composed of DB units by calculating the frequencies of their phonon modes. In Fig.~2(a) we provide the calculated phonon dispersions of TDS, LHDS, HDS and FDS showing that frequencies of all modes are positive over the whole Brillouin zone (BZ) for TDS, LHDS and HDS, while there are imaginary frequencies near the BZ boundary of FDS. This means that TDS, LHDS and HDS are thermodynamically stable structures while FDS is unstable. This also implies that the stability of TDS, LHDS and HDS structures does not depend on the substrate (i.e. Ag, in this case) and thus they can exist in their free-standing configuration. This is in contrast to the 3$\times$3 reconstructed silicene phase, which is dictated and preserved by the interaction of silicene with the silver substrate. This statement is further corroborated by the fact that the cohesive energy per atom of 3$\times$3 reconstructed silicene calculated in the absence of the Ag substrate is 0.109~eV lower than that of free-standing 1$\times$1 silicene.

\begin{figure}
\includegraphics[width=8.5cm]{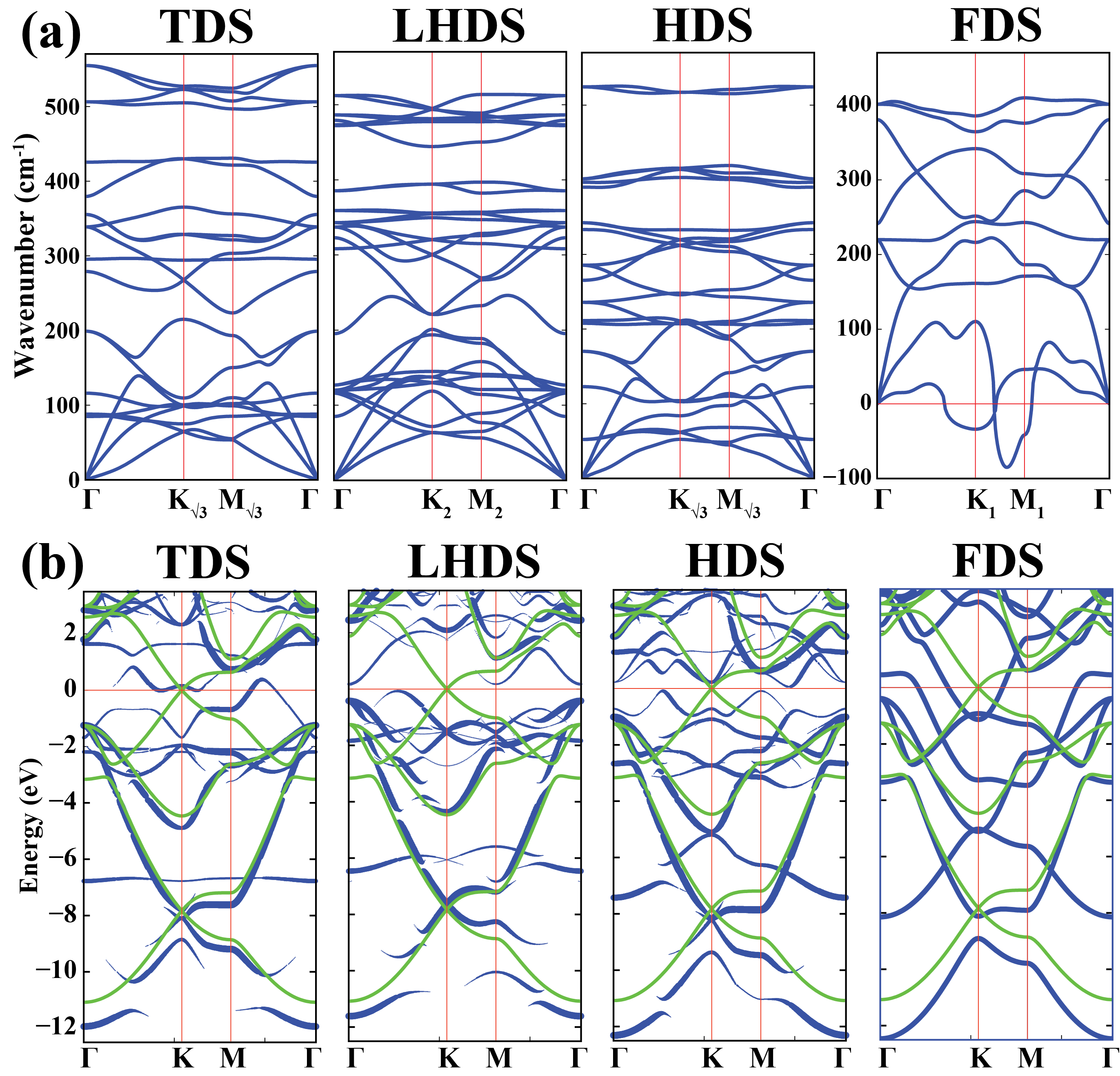}
\caption{(a) The phononic band dispersions of TDS, LHDS, HDS and FDS structures. The K and M points in the BZs of the 1$\times$1, 2$\times$2 and $\sqrt{3} \times \sqrt{3}$ unitcells are indicated by subscripts. (b) The electronic band dispersions of TDS, LHDS and HDS structures unfolded to the BZ of the free-standing 1$\times$1 silicene. The contribution from each state to the unfolded band is represented by a blue circle having a radius proportional to the weight of that state. The bands of FDS structure don't need to be unfolded because it has 1$\times$1 periodicity. The superimposed electronic band structures of the free-standing 1$\times$1 silicene are shown by green lines.}
\label{fig2}
\end{figure}

In Fig.~2(b) we present the electronic band dispersions of DB silicene structures. To compare with free-standing silicene we unfold the bands of all structures into the BZ of 1$\times$1 unitcell except that of FDS which already has this periodicity. We use the method described in Ref.~\cite{unfold} to do the unfolding \cite{hybridization}. The structures are intentionally ranked starting with TDS in which DB units are the most sparse and ending with FDS in which they are the most dense. This way one can immediately see how the flat band around -7~eV that comes from the weakly-interacting DB units of TDS is gradually turned into the highly-dispersive ($\sim$1.5~eV) band that comes from the strong interaction between DB units that are densely packed in the FDS structure. While this deep band of FDS is easily traced back to TDS the other band of FDS that is originating due to the DB units appears much higher and crosses the Fermi level. It is much harder to clearly associate this latter band with its counterparts in TDS, LHDS or HDS. This indicates that in these structures there is a complex interaction between the states originating from the DB units and the $\pi$-states coming from other Si atoms. These results could be used in experiments to identify the formation of DB units.

The growth mechanism of $\sqrt{3} \times \sqrt{3}$ silicene on Ag(111) substrate can be understood by analyzing the data presented in Table~I. Here we use five atomic layers to simulate the Ag(111) substrate. Note that, the 3$\times$3 silicene matches the 4$\times$4 Ag(111) supercell, while the DB structures can't be matched because their lattice constant is squeezed as the density of DB units are increased. To include the effect of Ag we first squeeze the 4$\times$4 Ag lattice to match the lattice of 3$\times$3 supercell of the DB structures and then optimize the system by keeping the Ag atoms fixed. Then we calculate the energy of squeezed Ag substrate in the absence of Si atoms. The energy difference between these two systems gives the cohesive energy of DB structures. Although 3$\times$3 structure has the lowest cohesive energy in the absence of Ag substrate, its cohesive energy surpasses that of cubic diamond silicon (4.598 eV/atom) when it is deposited on Ag(111) surface. This could be preventing Si atoms from clustering when they are deposited on Ag substrate. 

The DB units are spontaneously formed only if a monolayer silicene is already present. In this respect, the 3$\times$3 silicene acts as a precursor for the DB structures. When 3$\times$3 silicene is continuing to grow, the newly formed DBs diffuse and annihilate at the edges and contribute to the growth of 3$\times$3. Once 3$\times$3 silicene covers sufficiently large area, the DB units gradually organize themselves into the structure that has the largest cohesive energy. Here we look at the cohesive energy per area rather than cohesive energy per atom because the DB units compete to form the most energetic structure in the finite area covered by 3$\times$3 silicene. In this respect, the most favorable structure is $\sqrt{3} \times \sqrt{3}$ HDS \cite{vdw}. This process of growth is schematically summarized in Fig. 3(a). Here we note that since the HDS structure is incommensurate with Ag(111) surface it could be derived in the same fashion not only from 3$\times$3 reconstructed silicene but from any other monolayer phase of silicene on Ag \cite{eric}. This explains why the $\sqrt{3} \times \sqrt{3}$ reconstruction is usually observed at the advanced stages of silicene growth on Ag.

\begin{figure}
\includegraphics[width=8.5cm]{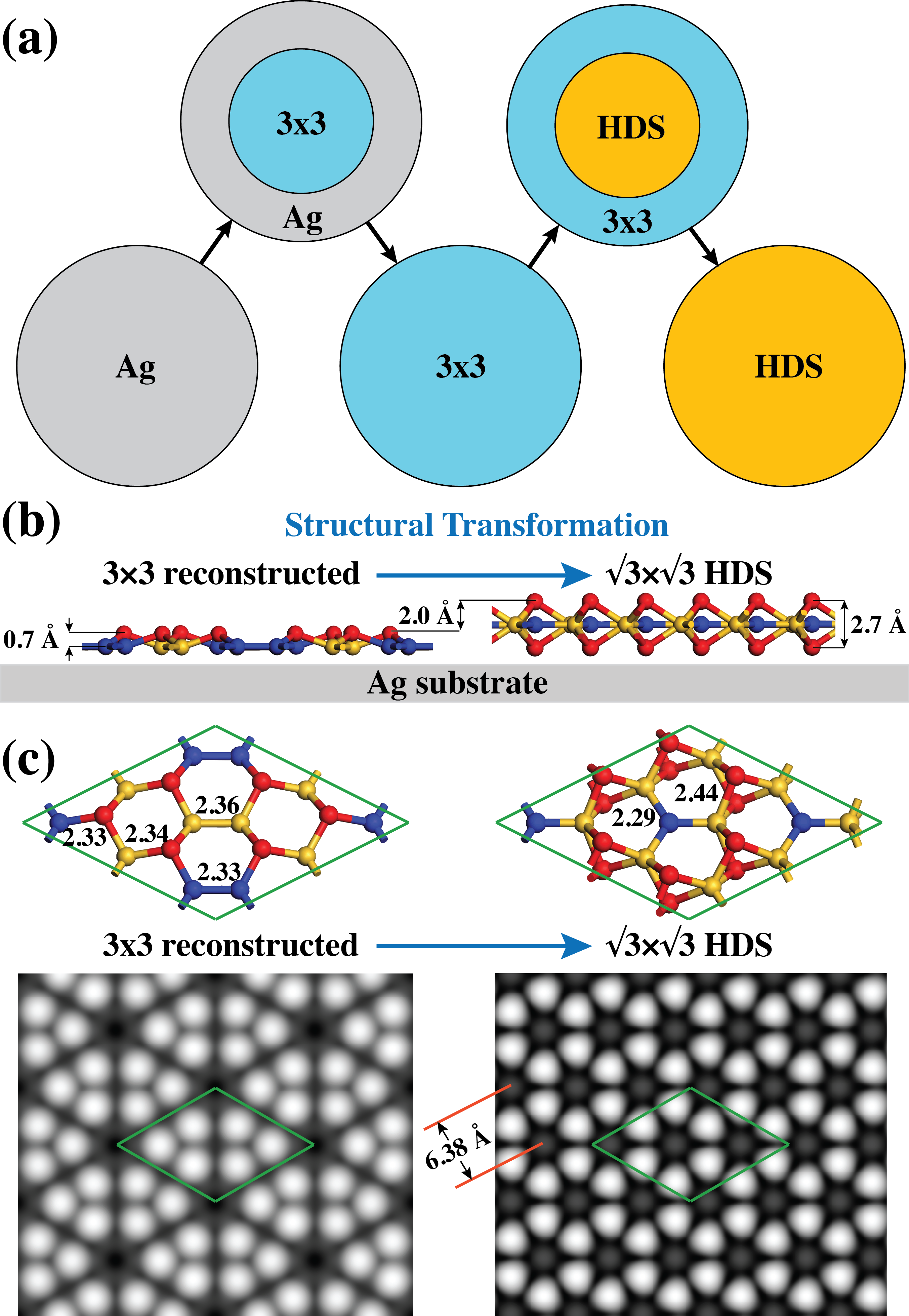}
\caption{(a) Growth of $\sqrt{3} \times \sqrt{3}$ HDS structure on Ag(111) surface. (b) Structural transformation from 3$\times$3 to $\sqrt{3} \times \sqrt{3}$ reconstructed silicene on Ag substrate. The height difference between two phases is approximately 2~\AA. (c) Ball and stick models and calculated STM images of 3$\times$3 reconstructed silicene and $\sqrt{3} \times \sqrt{3}$ HDS. Note that the bright spots in STM images coinside with protruding atoms shown by red balls. The numbers in the ball and stick model represent the bond lengths in~\AA. 3$\times$3 supercells are delineated by green lines.}
\label{fig3}
\end{figure}

Now we compare the calculated structural parameters of HDS with that of $\sqrt{3} \times \sqrt{3}$ phases derived from STM measurements. In the work by Chen \textit{et al.} an STM line profile going from Ag to silicene surface is reported providing an estimate for the vertical distance between the Ag and silicene surfaces to be 2.63~\AA \cite{chen}. This is in excellent agreement with our theoretical calculations for the thickness of the HDS structure, which we have calculated to be 2.66~\AA. Moreover, De Padova \textit{et al.} reported an STM line profile going from the 3$\times$3 to the $\sqrt{3} \times \sqrt{3}$ reconstructed silicene on Ag(111) surface \cite{padova}. In this case the vertical height difference between the two phases was reported to be approximately 2~\AA. As shown in Fig. 3(b), our calculation is in remarkable agreement also with this experimental result.

In Fig. 3(c) we present the structural details together with calculated STM images of the 3$\times$3 silicene and HDS. The former yields the well-known “flower pattern” while the honeycomb pattern seen in $\sqrt{3} \times \sqrt{3}$ HDS is very similar to the experimentally reported STM profiles \cite{feng,scirep,chen,lelay,eric}. Moreover, the calculated $\sqrt{3} \times \sqrt{3}$ lattice constant of HDS phase that yields 6.38~\AA~is in excellent agreement with the corresponding value deduced from the analysis of STM measurements that was reported to be 6.39~\AA~\cite{chen}. Note that, the lattice constant of HDS structure is $\sim$5~$\%$ shrunk compared to the lattice constant of the 3$\times$3 phase. This cannot be explained by the interaction of silicene and the Ag substrate as there is no way to commensurately match this lattice constant with Ag(111) surface. Instead, we propose that the structure of the HDS phase is intrinsic and is not dictated by the Ag surface (although the Ag(111) surface acts as a precursor of this structure through the stabilization of the 3x3 reconstruction for lower coverages) \cite{feng}. The fact that HDS is incommensurate with Ag(111) surface explains the experimentally observed Moire patterns in $\sqrt{3} \times \sqrt{3}$ phases of silicene \cite{feng,scirep,chen}.

\section{Conclusion}

In summary, we propose a growth model based on DB structures that form on silicene spontaneously and transform the 3$\times$3 phase into the $\sqrt{3} \times \sqrt{3}$ HDS phase. The calculated structural parameters of HDS is in excellent agreement with STM measurements. We show that, the HDS phase is intrinsic and incommensurate with Ag(111) surface. HDS is also stable in its free-standing form and its band structure is different from that of the free-standing silicene. Finally, we would like to mention that the intrinsic $\sqrt{3} \times \sqrt{3}$ structure of HDS could dictate the structure of multilayer silicene grown on top of it \cite{feng,padova,lelay}. Our findings could be extended to other group-IV elements like Ge and Sn.

\section{Acknowledgements}

SC and AR acknowledge financial support from the Marie Curie grant FP7-PEOPLE-2013-IEF project ID 628876, the European Research Council Advanced Grant DYNamo (ERC-2010-AdG-267374), Spanish Grant (FIS2010-21282-C02-01), Grupos Consolidados UPV/EHU del Gobierno Vasco (IT578-13), Ikerbasque and the European Commission projects CRONOS (Grant number 280879-2). VO\" O and SC acknowledge the support from TUBA. The Synchrotron SOLEIL is supported by the Centre National de la Recherche Scientifique (CNRS) and the Commissariat à l'Energie Atomique et aux Energies Alternatives (CEA), France.


\begin{thebibliography}{35}
\bibitem{vdw}
Including the van der Walls interaction does not change this conclusion, however the cohesive energies per area are increased for all silicene structures on Ag by $\sim$ 0.25 eV/\AA$^2$.
\end{thebibliography}
\end{document}